\documentclass[]{interact}

\usepackage{epstopdf}
\usepackage[caption=false]{subfig}
\usepackage{float}
\usepackage{xcolor}
\usepackage{algorithm}
\usepackage[natbibapa,nodoi]{apacite}
\usepackage{algpseudocode} 
\theoremstyle{plain}

\theoremstyle{definition}

\theoremstyle{remark}

\begin{document}

\articletype{Journal of Turbulence}

\title{Analysis of inertial-range intermittency in forward and inverse cascade  regions in isotropic turbulence}

\author{
\name{H. Yao\textsuperscript{a,b}\thanks{CONTACT H. Yao. Email: hyao12@jhu.edu} , 
   M. Schnaubelt\textsuperscript{a},
   A. Lubonja\textsuperscript{c},
   D. Medvedev\textsuperscript{a},
   Y. Hao\textsuperscript{a,b},
   M. Wang\textsuperscript{d},
   G. Lemson\textsuperscript{a},
   R. Burns\textsuperscript{a,c},
   A.S. Szalay\textsuperscript{a,c,g},
   P.K. Yeung \textsuperscript{e},
   G. Eyink\textsuperscript{a,f},
   T. A. Zaki\textsuperscript{a,b}, and 
   C. Meneveau\textsuperscript{a,b} 
}
\affil{\textsuperscript{a}Institute for Data Intensive Engineering and Science, Johns Hopkins University, Baltimore, MD, USA; \\
\textsuperscript{b}Department of Mechanical Engineering, Johns Hopkins University, Baltimore, MD, USA; \\
\textsuperscript{c}Department of Computer Science, Johns Hopkins University, Baltimore, MD, USA; \\
\textsuperscript{d}Department of Mechanical Engineering, Massachusetts Institute of Technology, Cambridge, MA, USA; \\
\textsuperscript{e}Schools of Aerospace Engineering and Mechanical Engineering, Georgia Institute of Technology, Atlanta, GA, USA; \\
\textsuperscript{f}Department of Applied Mathematics and Statistics, Johns Hopkins University, Baltimore, MD, USA;\\
\textsuperscript{g}Department of Physics and Astronomy, Johns Hopkins University, Baltimore, MD, USA}
}

\maketitle

\begin{abstract}
In order to test the hypothesis that inverse cascade regions in turbulent flows might exhibit more Gaussian noise-like and less intermittent small-scale statistics compared to the overall statistics, in this work we measure degrees of small-scale intermittency separately in regions of forward and inverse cascade. The local energy cascade rate $(\Phi_\ell)$ at length scale $(\ell)$ is defined using the scale-integrated Kolmogorov-Hill (KH) equation. The sign of $\Phi_\ell$ indicates the local cascading direction, with $\Phi_\ell > 0$ representing forward cascade of kinetic energy to smaller scales and $\Phi_\ell < 0$ representing inverse cascade from small to large scales. To characterize intermittency, we analyze the probability density functions (PDFs) of longitudinal and transverse velocity increments at scale $\ell$, conditioned on positive and negative $\Phi_\ell$. Our findings reveal that transverse velocity increments display approximately the same degree of non-Gaussianity and intermittency, regardless of whether they occur in forward or inverse cascade regions. The only noticeable difference is observed for longitudinal velocity increments that display strong negative skewness in regions of forward cascade compared to small positive skewness in regions of inverse cascade. We repeat the analysis for filtered velocity gradient tensor elements at scale $\ell$ and obtain similar results, except that the skewness of its longitudinal elements is slightly negative even in regions of inverse cascade. The result shows that subtle differences exist between unfiltered velocity increments at scale $\ell$ and gradients of filtered velocity, specifically in regions of inverse cascade where the additional small-scale information contained in velocity increments is necessary to establish inverse energy flux from small to large scales.  The analysis is based on isotropic turbulence data ($Re_\lambda \sim 1{,}250$) available from the public Johns Hopkins Turbulence Databases, JHTDB v2.0. This refactored system is based on the Zarr storage format, while data access is based on the ``virtual sensor'' approach, enabled by a Python backend package (Giverny) that replaces the legacy SQL storage and SOAP Web Services-based approaches. Information about the new system as well as sample Python notebooks are described and illustrated 
(Matlab, C, and Fortran access methods are also provided). 
\end{abstract}


\section{Introduction}

Turbulence is an inherently multi-scale phenomenon, whose main dynamical attribute is the 
energy cascade, i.e., on average, a transfer of turbulent kinetic energy from large to smaller scales, eventually down to viscous scales where it can be dissipated into heat. 
Quantitatively, the forward cascade is reflected in the  well-known $-4/5$ law \citep{kolmogorov1941local} relating the globally averaged third-order moment of  longitudinal velocity increments to the rate of viscous dissipation $\epsilon$. The relation reads $\langle \delta u_l(\ell)^3 \rangle = - \frac{4}{5} \ell \langle \epsilon \rangle$, where $\ell$ is a length within the inertial range and $\delta u_l(\ell)$ is the velocity difference over a distance $\ell$ in the direction of the velocity difference vector.

Additionally, turbulent flows are known to be highly intermittent \citep{frisch1995turbulence}, and as a result, averaged quantities such as the mean dissipation rate $\langle \epsilon \rangle$ cannot fully describe many highly relevant statistics of turbulent fluctuations. 
For example, the dissipation rate is highly intermittent \citep{kolmogorov1962refinement,meneveau1991multifractal, sreenivasan1997phenomenology} leading to velocity structure function scalings \citep{anselmet1984high} that deviate from the original predictions of Kolmogorov \citep{kolmogorov1941local}. Kolmogorov then proposed a more detailed relationship between velocity increments and dissipation, stating that the statistics of velocity increments are determined by the local spherically averaged dissipation rate (at scales in the inertial range) rather than by the globally averaged dissipation \citep{kolmogorov1962refinement}. This ``refined similarity hypothesis'' (KRSH) has been validated experimentally \citep{praskovsky1992experimental,stolovitzky1992kolmogorov, lawson2019direct} and numerically  \citep{wang1996examination,iyer2015refined, yeung2020advancing}. A recent analysis of this hypothesis \citep{yao2024forward} demonstrated the validity of this hypothesis for local, spherically integrated structure functions, which allowed a more direct connection to the Navier-Stokes equations written at two points. This type of local spherically integrated structure function arises from the generalized Kolmogorov-Hill equation (also known as Karman-Howarth-Monin-Hill KHMH equation \citep{yao_papadakis_2023, portela_papadakis_vassilicos_2017_2}), integrated over a sphere of diameter $\ell$. It was argued in \cite{yao2024comparing} that this quantity, denoted as $\Phi_\ell$, represents a true measure of local cascade rate of energy between scales since it arises from the volume integral of divergence in scale-space.  Using this formulation, one can identify local inverse or forward cascade unambiguously, specifically  $\Phi_\ell > 0$ denoting energy flux to smaller scales  (local forward cascade), while $\Phi_\ell < 0$ indicates  flux from small to large scales, i.e. local inverse cascading of energy. 

In the analysis of \cite{yao2024comparing}, the focus was on finding relationships between local forward/inverse energy cascade rate at scale $\ell$ and large-scale (inertial range) local flow deformation and rotation rates at the same scale. The statistics of the filtered velocity gradient tensor ($\widetilde{A}_{ij} = \partial \widetilde{u}_i/\partial x_j$, where $\widetilde{u}_i$ represents the velocity component in the $x_i$ direction, filtered at scale $\ell$) are closely linked to turbulence intermittency, playing a crucial role in the energy cascade \citep{borue1998local,van2002effects,meneveau2011lagrangian, johnson2024multiscale}. Specifically, in \cite{yao2024comparing}, averages of $\Phi_\ell$ conditioned on the invariants of the local filtered velocity gradient tensor, i.e., $Q$ and $R$, were examined, and the statistics of $Q$ and $R$ were quantified depending on whether $\Phi_\ell>0$ or $\Phi_\ell<0$. It was observed that the joint PDFs of $Q$ and $R$ exhibited significant differences between forward and inverse cascade regions.
For the forward cascade regions, the joint PDF of $(R,Q)$ displayed the typical ``tear-shape" asymmetric pattern \citep{cantwell1993behavior} also observed in many turbulent flows across scales \citep{meneveau2011lagrangian}. Conversely, the joint PDF of $(R,Q)$ within the inverse cascade regions demonstrated a notable left-right symmetry, resembling the joint PDFs that result from a Gaussian velocity field. 

This observation raises the possibility that local inverse cascade regions in turbulence might behave in a more Gaussian, non-intermittent fashion. We recall that Large Eddy Simulation (LES) models that include stochastic terms (to model backscatter, i.e. inverse cascading), are based on adding random noise terms with Gaussian statistics to the filtered Navier-Stokes (LES)  equations \citep{leith1990stochastic,mason1992stochastic,piomelli1991subgrid}. Hence we are interested in quantifying the degree of intermittency and non-Gaussianity separately in regions of forward and inverse cascade, in order to establish if, in the latter, Gaussian, more random behavior dominates.  The most common object that has been studied to detect intermittent turbulence statistics at some scale $\ell$ are the longitudinal and transverse velocity increments (denoted here as $\delta u_l(\ell)$ and $\delta u_t(\ell)$, respectively), whose moments are called structure functions \citep{anselmet1984high,shen2002longitudinal,chevillard2006unified}. We shall quantify the statistics of $\delta u_l(\ell)$ and $\delta u_t(\ell)$ separately in regions of forward and inverse cascade. In addition to velocity increments, we will analyze the statistics of the filtered velocity gradient tensor $\widetilde{A}_{ij}$\citep{borue1998local,van2002effects,meneveau2011lagrangian, johnson2024multiscale} at various scales $\ell$, again separating the statistics into forward and inverse cascading regions. 

To calculate the local energy cascade rate $\Phi_\ell$, as well as  longitudinal and transverse velocity increments and components of the filtered velocity gradient tensor, we use data from DNS of forced isotropic turbulence at a Taylor-scale Reynolds number of $Re_\lambda \sim 1{,}250$, computed on a computational grid of $8{,}192^3$ points. 
Relevant background on the definition of $\Phi_\ell$ and its interpretation as a local energy cascade rate is provided in \S \ref{sec:scaleKHeq}. Results concerning velocity increments in the inertial range are presented in \S \ref{sec:results1}. Results concerning the levels of intermittency of filtered velocity gradient elements in the inertial range are presented in \S \ref{sec:results2}.  Conclusions are summarized in \S \ref{sec:conclusions}.

The data to be used in the present analysis are available from the Johns Hopkins Turbulence Database (JHTDB) system, which has been overhauled to increase the number of available datasets and enhance the robustness of data access tools.   The new framework (JHTDB 2.0) is based on a Python  backend code package that replaces  the previous one based on C\#. For the new system (\url{https://turbulence.idies.jhu.edu/home}), a single function {\it getData} enables access to all available datasets using the virtual sensor method (users send arrays of points and times to the database system and the latter returns data at those points and times). The new {\it getData} function unifies legacy turbulence services including queries for local velocity, pressure, and their first and second derivatives, enabling users to obtain these data using a single function call.  Datasets include DNS of homogeneous isotropic turbulence at various Reynolds numbers, several wall-bounded turbulent flows, LES of stratified atmospheric boundary layer flow and windfarms, etc. Data files are stored on a Ceph-FS cluster using the Zarr storage method, which subdivides the data arrays into small 3D chunks whose size has been optimized. Data access is possible using Python or Matlab notebooks, as well as C and Fortran codes. These can be run close to the data on a dedicated server (SciServer) or on a user's platform with data transfer made possible using the representational state transfer (REST) interface method. 
Additional details about the refurbished JHTDB system are provided in the appendix.

\section{Quantifying local cascade rates with the scale-integrated KH equation}
\label{sec:scaleKHeq}

The scale-integrated KH equation \citep{yao2023entropy,yao2024comparing} is an exact transport equation for the evolution of local second-order velocity increments \citep{hill2001equations, hill2002exact}, integrated over a sphere in scale (${\bf r}$)-space, extending up to a scale of size $\ell$ (or sphere radius of $\ell/2$). We use ${\bf r}$ to represent the vector separating two points over which the velocity increment is defined, and we use ${\bf r}_s = {\bf r}/2$ as the radial coordinate vector originating from the local reference point denoted as ${\bf x}$, i.e. the center of the local sphere. The local structure function-based kinetic energy of turbulence at all scales smaller or equal to $\ell$ can be defined according to:

\begin{equation}
k_{\ell}({\bf x},t) = \frac{1}{2 \,V_\ell}\iiint\limits_{V_{\ell}}  \frac{1}{2} \delta u _i^2({\bf x},{\bf r}) \, d^3{\bf r}_s,
\label{eq:defk}
\end{equation}
where $ \delta u_i({\bf x},{\bf r}) =  u_i({\bf x}^+,t) - u_i({\bf x}^-,t)$ is the velocity increment vector in the $i^{\textrm{th}}$ Cartesian direction between two points ${\bf x}^+={\bf x}+{\bf r}/2$ and ${\bf x}^-={\bf x}-{\bf r}/2$, separated by ${\bf r} = {\bf x}^+ - {\bf x}^-$ and middle point 
${\bf x} = ({\bf x}^+ + {\bf x}^-)/2$.  The integration in Eq. \ref{eq:defk} is performed over a ball with volume $V_\ell=\frac{4}{3}\pi( {\ell}/{2})^3$ with a diameter equal to $\ell$. 
The factor 2 in the denominator in front of the integral is to avoid double counting the energy content since the integration over the entire sphere counts the same energy upon exchanging ${\bf x}^+$ and  ${\bf x}^-$. In its instantaneous form, and ignoring any forcing term, the dynamical evolution equation for $k_{\ell}({\bf x},t)$ can be derived from the Navier-Stokes equation at two points \citep{hill2001equations, hill2002exact,yao2024comparing}, and is expressed as follows:

\begin{equation}
 \begin{aligned}
{\frac{\widehat d k_\ell}{dt}}
=
\Phi_\ell
-
 \epsilon_\ell
-
P_\ell
+
D_\ell
.
\end{aligned}
\label{ins_GKHE_noint}
\end{equation}

The various terms and their physical meanings are the following, beginning with the first term on the right-hand side of the equation:

(1) The energy cascade rate at the length scale $\ell$:
\begin{equation}
\Phi_\ell  \equiv -\frac{3}{4\,\ell}\frac{1}{S_\ell}\oint\limits_{S_{\ell}} \delta u _i^2\,\delta u _j\,  \hat{n}_j \, dS,
\label{phiflux}
\end{equation} where ${S_\ell} = 4\pi(\ell/2)^2$ represents the area of the sphere with a diameter equal to $\ell$, and $\hat{n}_j$ is the unit normal vector pointing outward from the sphere. Gauss’ theorem in scale space has been applied to the divergence term of the third-order structure function term. The sign of $\Phi_\ell$ serves as an indicator of the energy flux direction within  scale space at position ${\bf x}$ within eddies of length scale $\ell$. When $\Phi_\ell>0$ the cascade is forward and energy flux leads to an increase of small-scale structure function-based kinetic energy, and vice-versa. In this paper, it is this quantity that will be used to determine if locally the cascade of kinetic energy is forward or inverse. As discussed in \cite{yao2023entropy,yao2024comparing} the definition of $\Phi_\ell$ represents a genuine flow of kinetic energy between scales, since its definition is based on a flux vector field in ${\bf r}$ scale space.  

(2) Viscous dissipation:

\begin{equation}
\epsilon_\ell \equiv \frac{1}{V_\ell}\int\limits_{V_{\ell}}
 \epsilon^*({\bf x},{\bf r}) \, d^3{\bf r}_s,
\end{equation}
where the symbol $\epsilon$ represents the ``pseudo-dissipation'', defined as $\epsilon =\nu ({\partial u_i}/{\partial x_j})^2$, and  $\epsilon^*({\bf x},{\bf r})=  (\epsilon^+ + \epsilon^-)/2$, where we utilize the superscript $*$ to denote the average value between two points (not the value at the center). 

(3) Pressure and velocity increments correlation:

\begin{equation}
P_\ell  \equiv -\frac{6}{\ell}\frac{1}{S_\ell}\oint\limits_{S_{\ell}} \frac{1}{\rho} \, p^* \, \delta u _j \, \hat{n}_j \,dS.
\end{equation}

(4) Viscous diffusion in physical and length scale spaces:

\begin{equation}
 \begin{aligned}
D_\ell \equiv 
\frac{\nu}{4}\frac{1}{V_\ell}\iiint\limits_{V_{\ell}}\left(\frac{1}{2} \frac{\partial^2 \delta u_i^2}{\partial x_j \partial x_j 
}
+
2 \frac{\partial^2 \delta u_i^2}{\partial r_j \partial r_j}
\right)
d^3{\bf r}_s.
\end{aligned}
\label{Dell}
\end{equation}

(5) On the left-hand side is the rate of change of kinetic energy at all scales smaller or equal to $\ell$:
\begin{equation}
 \frac{\widehat{d} k_\ell}{dt} \equiv  \frac{\partial k_\ell}{\partial t} + 
 \frac{1}{2 \,V_r}\int\limits_{V_{r}}   u^*_{j} \, \frac{\partial \frac{1}{2}\delta u _i^2}{\partial x_j}   \, d^3{\bf r}_s,
\end{equation}
where the two-point averaged velocity is defined as $u^*_j = (u_j^+ + u_j^-)/2$. The second term represents spatial advective transport and includes advection by larger scales but also smaller-scale turbulent transport \citep{yao2023entropy}. 

The local dissipation $\epsilon_\ell$ is directly relevant to Kolmogorov's Refined Similarity Hypothesis (KRSH). It was shown in \citep{yao2024forward} that when evaluating conditional averages of Eq. \ref{ins_GKHE_noint} based on $\epsilon_\ell$, the conditional averages of the unsteady, pressure, and viscous terms are typically negligible, and therefore the only terms that remain are
$\langle \Phi_\ell | \epsilon_\ell\rangle \approx \langle \epsilon_\ell | \epsilon_\ell\rangle=\epsilon_\ell$, i.e. a version of KRSH that therefore can be directly connected to the Navier-Stokes equations written at two points via the scale-integrated local KH equation \citep{yao2024forward}. 
In the present study, we will focus on $\Phi_\ell$ and $\epsilon_\ell$ in Eq. \ref{ins_GKHE_noint}, while other terms will not be the focus in this study.

With a clear definition of energy cascade rate $\Phi_\ell$ directly connected to the evolution of kinetic energy at and below scales $\ell$ via Eq. \ref{ins_GKHE_noint}, we can proceed to quantify the level of intermittency separately in regions of forward and inverse cascade. To characterize intermittency in these distinct flow regions, we measure statistics of longitudinal and transverse velocity increments at scale $\ell$, and velocity gradients across separation distance $\ell$. The latter has been studied extensively in the past \citep{borue1998local,van2002effects,meneveau2011lagrangian} since it encapsulates rich multiscale information not only about intermittency but also about the local fluid deformation, rotation, and their directions at arbitrary length-scale $\ell$. Past research on statistics of $\widetilde{A}_{ij}$ has shown that its tensor elements display strong non-Gaussian  statistics and wide tails in their PDFs, with increasing levels of intermittency as $\ell$ decreases \citep{chevillard2006lagrangian}. Its longitudinal elements $\widetilde{A}_{ll}$ (no summation over indices $l$) display negative skewness in addition to wide tails and flatness factors exceeding the Gaussian value of 3, while the transverse components (denoted as $\widetilde{A}_{tt}$ (again no summation over `$t$' and indicating $\widetilde{A}_{pq}$ with $p\neq q$) display symmetric PDFs (zero skewness) but even larger flatness factors \citep{meneveau2011lagrangian}. 

For consistency with the spherical volume integration used in establishing Eq. \ref{ins_GKHE_noint},  we apply a spherical top-hat filter to compute the filtered velocity according to
\begin{equation}
\widetilde{u}_i ({\bf x},t) = \iiint G_\ell({\bf r}_s) \, {u_i}({\bf x + r}_s,t) \, d {\bf r_s}
\end{equation}
where $G({\bf r})$ represents the filter kernel characterized by the length scale $\ell$, and ${\bf r_s}$ is the radius vector. For the top-hat filter used here $G_\ell({\bf r}_s) = 1/V_\ell$ for $|{\bf r}_s| \leq  \ell/2$ and zero  otherwise.  The results presented in this paper focus on a scale in the inertial range, $\ell = 45 \eta$, (where $\eta = (\nu^3 / \langle \epsilon \rangle)^{1/4}$ is the Kolmogorov length scale, and $\langle \epsilon \rangle$ is the globally averaged dissipation) while some other scales ($\ell = (30,60,75) \eta$) are considered as well to establish relative independence of main results as function of scale.   
 
\begin{figure} [H]
 \centering
  \includegraphics[scale=0.5]{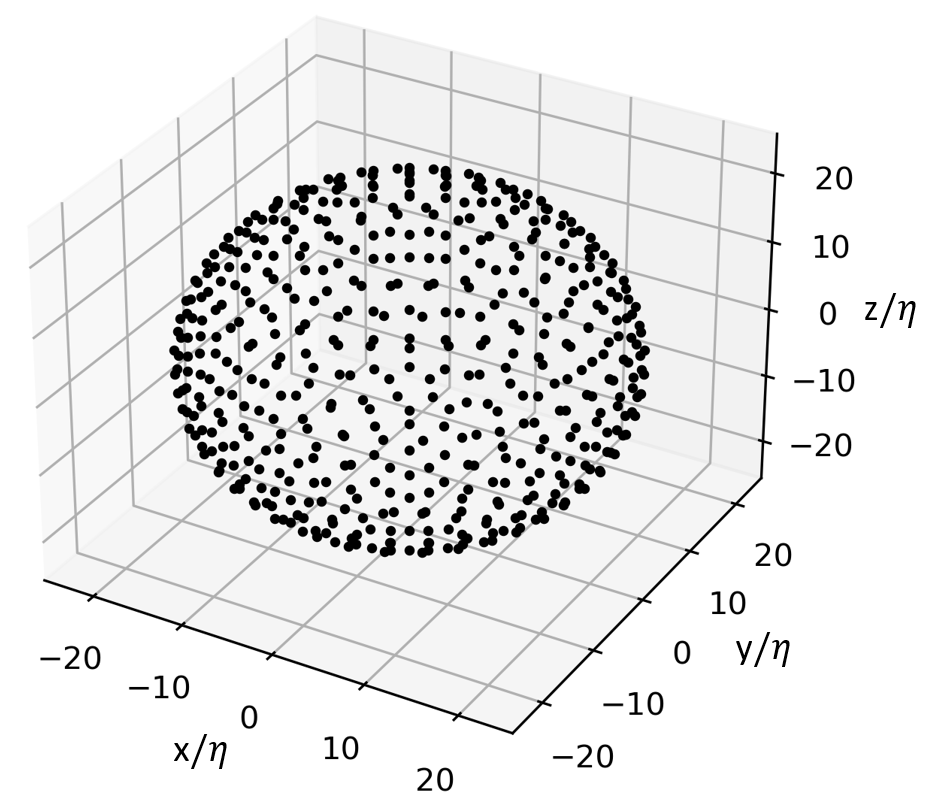}
    \caption{Generation of 500 uniformly distributed points on the surface of sphere with $\ell = 45\eta$.}
    \label{fig: sphere}
\end{figure}

In this study, we analyze data from Direct Numerical Simulation (DNS) of forced isotropic turbulence at $Re_\lambda = 1{,}250$ \citep{yeung2015extreme} where $Re_\lambda$ represents the Taylor-scale Reynolds number. The DNS dataset is stored using the Zarr format on a Ceph-FS cluster, and the data are publicly accessible via the recently refurbished and updated JHTDB v2.0 (Johns Hopkins Turbulence Database) data access tools. More details about data storage are provided in the appendix. The data are used to calculate local energy cascade rate $\Phi_\ell$ and its sign is used to classify a point at the center of the sphere as either forward or inverse cascade point. To compute $\Phi_\ell$, velocities at many pairs of points in an ensemble of spheres of diameter $\ell$ must be accessed efficiently. This is accomplished using the {\it virtual sensors} approach that forms the basis of the JHTDB data access philosophy, instantiated in the new {\it getData} function. For details about this tool, see the appendix. For each sphere, we use a set of $N_p$ pairs of points, distributed approximately uniformly on the sphere. For the baseline case of $\ell = 45 \eta$, we use $N_p=500$ points, yielding an average distance between sampling points of about $3.6\eta$ ($N_p=2000$ points are utilized for the largest $\ell/\eta=75$ case). Sensitivity tests using finer samplings \citep{yao2022analysis} show indistinguishable results. The points are equidistributed points on the surface of each sphere using the algorithm by \cite{deserno2004generate}. Fig. \ref{fig: sphere} shows one sphere with $\ell = 45 \eta$ and 500 points on the surface distributed uniformly.

Since these points do not have to lie on DNS grid-points, spatial interpolation is used. The {\it getData} function allows users to specify among several types of Lagrange polynomial or spline spatial interpolation methods. Here we specify 8th-order Lagrange spatial interpolation to obtain velocity vectors at the $N_p$ points on each of the spheres to compute $\Phi_\ell$ as well as transverse and longitudinal velocity increments. Then, statistics are collected over $N_s = 2\times 10^6$ spheres whose center positions ${\bf x}$ are selected randomly (uniform spatial distribution) within the full $(2\pi)^3$ DNS domain.  The pseudocode selecting $N_p$ points on a surface and then a sample over $N_s$ spheres is shown in the pseudocode below. For each sphere, $N_s$ pairs of points ($\pm$) are assembled and velocity values are queried using the {\it getData} function.

\begin{algorithm}
\vspace{0.2in}
    \caption{\bf Pseudocode: Velocity sampling at $2N_p$ points on $N_s$ spheres using getData}
    \begin{algorithmic}[1]
        \State Step (1): Assemble array ${\bf P}_{\text{surface}}$ of coordinates for $N_p$ points, equi-spaced on sphere of radius $R$.
        \State \noindent\rule{\linewidth}{0.4pt}
        \State $\circ$  Define sphere radius: $R \gets \frac{45}{2}\eta$
        \State $\circ$ Select $N_p$ pairs of latitude and azimuthal angles $(\nu, \phi)$ to have approximately constant distributed points on a sphere \citep{deserno2004generate}
        \For{each of the $N_p$ pairs  $(\nu, \phi)$:}
                \State \hskip1em $x_p \gets R \sin(\nu) \cos(\phi)$
                \State \hskip1em $y_p \gets R \sin(\nu) \sin(\phi)$
                \State \hskip1em $z_p \gets R \cos(\nu)$
                \State Append $(x_p, y_p, z_p)$ to ${\bf P}_{\text{surface}}$
       \EndFor \\ 
       \State(see Fig. 1 illustrating $N_p=500$ points on a sphere of diameter 45$\eta$)
       \State \noindent\rule{\linewidth}{0.4pt}
        
       \State Step (2):  For $N_s$ spheres, read velocity at $2N_p$ points on sphere
       \State \noindent\rule{\linewidth}{0.4pt} 
       \State $\circ$  Generate $N_s$ random points $(x_c, y_c, z_c)$ uniformly in $[0, 2\pi]^3$
            \State $\circ$  Loop over random points (sphere centers):
        \For{each sphere $i = 1$ to $N_s$}
            \State Set center point: $\mathbf{c}_i \gets (x_c, y_c, z_c)$
            \State Compute $2N_p$ physical locations on the shell:
            \State ${ points}^{(i)}  \gets \mathbf{c}_i + {\bf P}_{\text{surface}}$
            \State  Append $\mathbf{c}_i - {\bf P}_{\text{surface}}$ to ${ points}^{(i)}$  
            \State Query velocities at those $2N_p$ points:
        \State \hskip1em $velocities^{(i)}  \gets \texttt{getData}(\texttt{"isotropic8192"}, \texttt{"velocity"}, \texttt{"field"},$
        \State \hskip7em \hskip5em ${\text{points}}^{(i)}, \ldots)$

        \EndFor

        \State \noindent\rule{\linewidth}{0.4pt}    
    \end{algorithmic} 
    \vspace{0.2in}
\end{algorithm}

To compute the filtered velocity gradients $\widetilde{A}_{ij}$, the refurbished  {\it getCutout} function (see appendix) is used. For any given ${\bf x}$ (center of the sphere) we invoke the cutout function for a cube of data with dimensions of $\ell^3$ centered at ${\bf x}$. The extracted cube of data is then multiplied by a spherical top-hat kernel function (a mask with zeroes at distances greater than $\ell/2$ from the cube's center) to obtain the local filtered velocity vector $\widetilde{u}_i({\bf x})$. Using the same filtering method, we also measure filtered velocities at 2 grid-points on both sides of the center points in each Cartesian direction. Then, a fourth-order central difference scheme is applied to calculate the filtered velocity gradient tensor elements corresponding to position $(\bf x)$.
 
\section{Velocity increment statistics in forward and inverse
cascade regions}
\label{sec:results1}

In this section, we report the measured PDFs of longitudinal ($\delta u_{l}(\ell)$) and transverse ($\delta u_{t}(\ell)$) velocity increments, for $\ell = 45 \eta$. At each point, three pairs of $+$ and $-$ points are chosen in the Cartesian directions. Therefore the samples of longitudinal increments include $\delta u_{1}=u_1({\bf x}+{\bf r}/2) - u_1({\bf x}-{\bf r}/2)$ with ${\bf r} = \ell \, {\bf i}$, and similarly $\delta u_{2}$ with ${\bf r} = \ell  \, {\bf j}$ and $\delta u_{3}$ with ${\bf r} = \ell \, {\bf k}$, i.e., 6 million samples of longitudinal velocity increments for 2 million samples of energy cascade rate $\Phi_\ell({\bf x})$. The samples of transverse increments $\delta u_{t}$ include $\delta u_{2}$ and $\delta u_3$ for ${\bf r} = \ell \, {\bf i}$, etc. for a total of 12 million samples of transverse velocity increments. Three PDFs are shown: one conditioned on local forward cascade (when $\Phi_\ell({\bf x}) > 0$), another for inverse cascade points ($\Phi_\ell ({\bf x})  < 0$), and the third the overall unconditioned statistics, using the full sample of 2 million randomly distributed points across the entire domain.  Observations about  PDFs are made quantitative using higher moments of the velocity increments, namely skewness and flatness, defined as follows,
\begin{equation}
 \begin{aligned}
S \equiv \frac{ \langle \delta u^3 \rangle}{\langle \delta u^2 \rangle^{(3/2)}}, \quad F \equiv \frac{ \langle \delta u^4 \rangle}{\langle \delta u^2 \rangle^2},
\end{aligned}
\label{skewandFlat}
\end{equation}
where the averaging is either global averaging over the entire ensemble or using conditional averaging as indicated. 

\begin{figure} [H]
 \centering
  \includegraphics[scale=0.5]{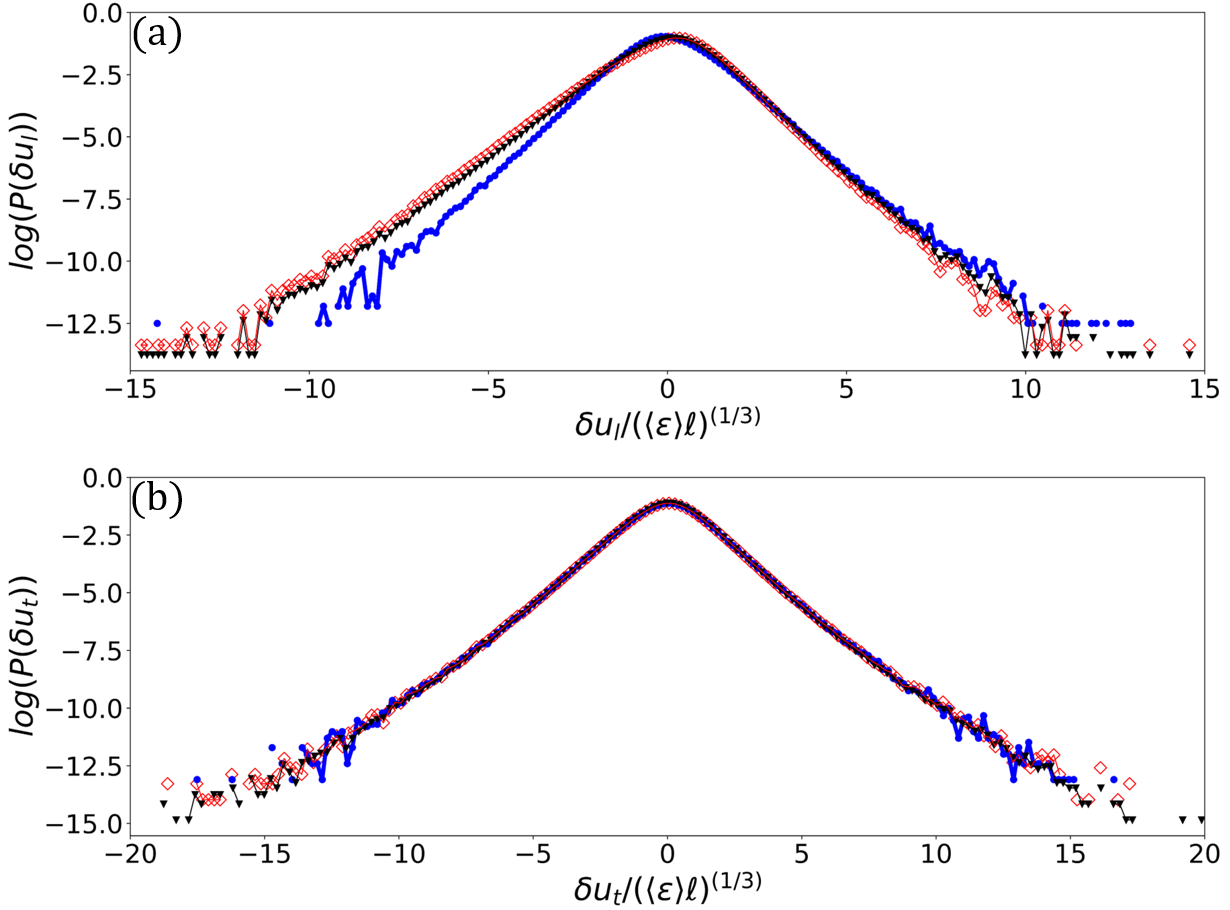}
    \caption{PDFs of velocity increments in isotropic turbulence at scale at $\ell = 45\eta$ for:  global conditions including both forward and inverse cascade regions (black triangles and line), forward cascade regions (red open diamonds and line) and inverse cascade regions (blue circles and line). Panel (a) is for longitudinal increments while panel (b) is for transverse velocity increments. In this and all other PDFs, natural logarithm is used.}
    \label{fig: PDFs_45eta_du}
\end{figure}

The measured PDFs are shown in Figure \ref{fig: PDFs_45eta_du}. Panel (a) depicts the PDFs of $\delta u_{l}$. The unconditional (global) PDFs ($P(\delta u_{l})$ shown as black triangles and lines) display the well-known elongated stretched exponential tails (flatness $F=4.99$), as well as the negative skewness ($S = -0.29$, a value consistent with the -4/5 law and a Kolmogorov constant $C_2 \approx 2$, since $S = -4/5 \, C_2^{-3/2}$). Panel (b) shows the PFDs of the transverse increments, $P(\delta u_{t})$, displaying even longer tails (flatness $F=5.66$) and no skewness ($S\approx 0$), the well-known behavior of velocity increments. Results for conditioning on forward cascade ($P(\delta u_{l}|\Phi_\ell>0)$ shown as red diamonds and lines), and conditioning on inverse cascade ($P(\delta u_{l}|\Phi_\ell<0)$ shown as  blue circles and lines) are also displayed in Fig. \ref{fig: PDFs_45eta_du} (flatness of $F \approx 4.97$ and $4.93$ respectively).  The transverse components shown in (b) yield  the same PDFs to within statistical accuracy (all have flatness of $F \approx 5.6-5.7$), providing direct evidence that turbulence in both forward and inverse cascade regions display very much the same levels of intermittency. The longitudinal increments show expected behavior: the skewness of the PDFs is slightly more pronounced in forward cascading regions ($S = -0.46$) while in regions of inverse cascade the skewness is slightly positive ($S = 0.17$).

\begin{figure} [H]
 \centering
  \includegraphics[scale=0.45]{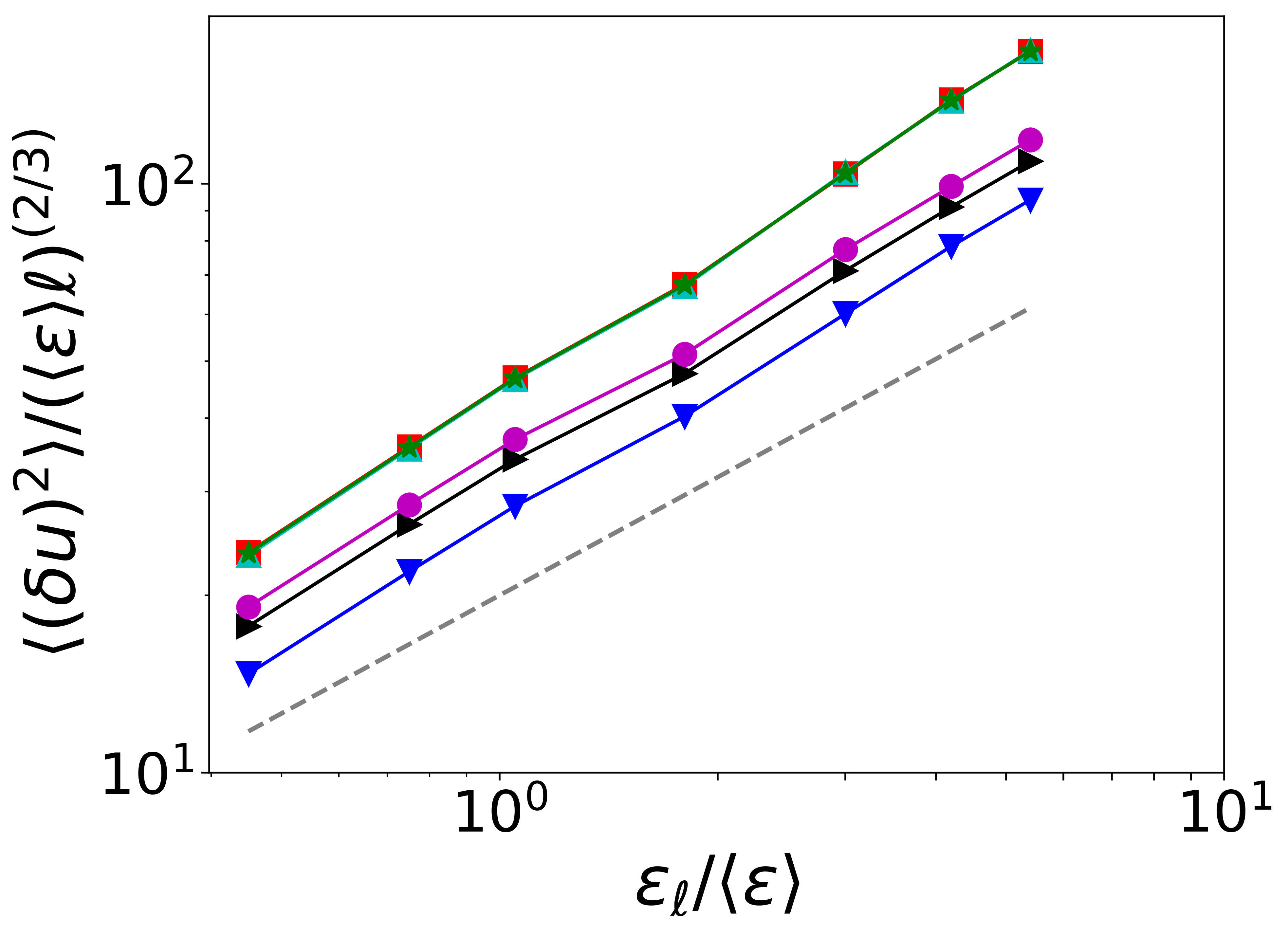}
    \caption{Averaged second moment of velocity increments conditioning on local dissipation and forward or inverse cascade rates: $((\delta u_{l})^2 | \epsilon_\ell,\Phi_\ell > 0)$ is marked in purple line and circles, $((\delta u_{t})^2 | \epsilon_\ell,\Phi_\ell > 0)$ is marked in red line and squares), $((\delta u_{l})^2 | \epsilon_\ell,\Phi_\ell < 0)$ is marked in blue downwards pointing triangles , and $((\delta u_{t})^2 | \epsilon_\ell,\Phi_\ell < 0)$ is marked in cyan upward pointing triangles. Results without conditioning on the sign of $\Phi_\ell$ are shown as black triangle and dark green star for the longitudinal and transverse components, respectively. The grey dashed line has a slope of $2/3$. In this plot, the base of the logarithm is 10.}
    \label{du2_45eta_eps}
\end{figure}

Next, in the spirit of the KRSH \citep{kolmogorov1962refinement,stolovitzky1992kolmogorov} and the recent analysis of \cite{yao2024forward}, we compute statistics of velocity increments conditioning on both local viscous dissipation rate $\epsilon_\ell$, as well as on the sign of energy cascade rate across scales. We begin by evaluating second-order moments and present results as a function of local dissipation $\epsilon_\ell$ in Fig. \ref{du2_45eta_eps}. In general, KRSH is predicts a 2/3 power-law scaling $\langle \delta u^2\rangle \sim \epsilon_\ell^{2/3} \, \ell^{2/3}$. As shown in Fig. \ref{du2_45eta_eps} this KRSH scaling holds irrespective of whether conditioning on forward or inverse cascading regions. Interestingly, the magnitude of the longitudinal velocity increments (conditional second order structure functions) depends on the local cascade rate, with the inverse cascade associated with the smallest magnitude velocity increments, perhaps indicative of a depletion of energy at scale $\ell$ because it is being transferred to larger scales. 
Conversely, the transverse component is entirely oblivious to the direction of the local energy cascade rate with all three conditional averages (global, positive and negative $\Phi_\ell$) indistinguishable. 

\begin{figure} [H]
 \centering
  \includegraphics[scale=0.65]{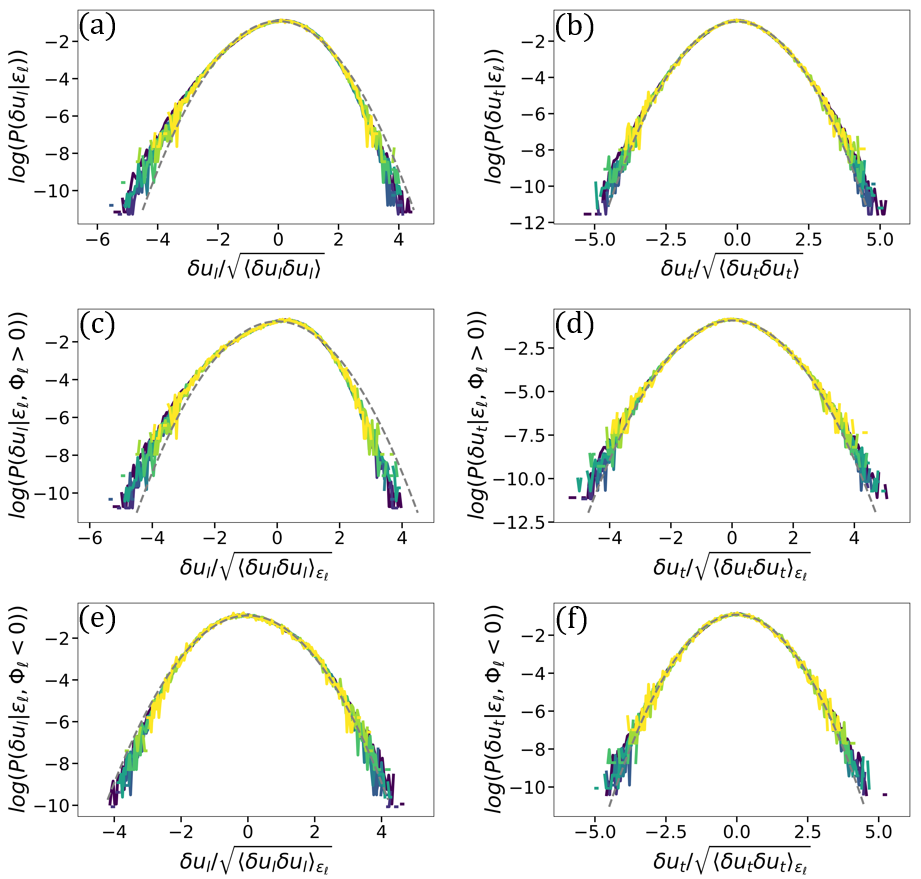}
    \caption{PDFs of velocity increments conditioning on local viscous dissipation rate (conditioned on ranges in bins centered at $\epsilon_\ell / \langle \epsilon \rangle = 0.15$ (black), 0.45, 0.75, 1.05, 1.8, 3.0, 4.2, and 5.4 (yellow)) and forward or inverse cascade rates ($\Phi_\ell > 0 $ or $\Phi_\ell < 0$). Panel (a): PDFs of $ \delta u_{l}$ conditioned on $\epsilon_\ell$, panel (b): Conditional PDFs of $\delta u_{t}$, panel (c): Conditional PDFs of $\delta u_{l}$ conditioned on $\epsilon_\ell$ and $\Phi_\ell>0$, panel (d): Conditional PDFs of $\delta u_{t}$ conditioned on $\epsilon_\ell$ and $\Phi_\ell>0$, panel (e): Conditional PDFs of $\delta u_{l}$ conditioned on $\epsilon_\ell$ and $\Phi_\ell<0$, Panel (f): Conditional PDFs of $\delta u_{t}$ conditioned on $\epsilon_\ell$ and $\Phi_\ell<0$. The grey dashed line is a Gaussian distribution with zero mean and unit standard derivation.}
    \label{fig:PDFs_45eta_du_eps}
\end{figure}

We then study the conditional PDFs for different $\Phi_\ell$
and $\epsilon_\ell$. A remarkable result from prior research \citep{stolovitzky1992kolmogorov,chevillard2006unified} related to the KRSH is that the statistics of $\epsilon_\ell$-conditioned velocity increments in the inertial range become non-intermittent and much closer to Gaussian than the unconditional values. That is to say, the level of intermittency of turbulence appears to be entirely encoded in the statistics of $\epsilon_\ell$ which become more and more intermittent with decreasing scale and increasing Reynolds number \citep{meneveau1991multifractal,frisch1995turbulence}. We here explore this concept by further conditioning on the sign of the energy cascade rate. Results are shown in Figs. \ref{fig:PDFs_45eta_du_eps}. Remarkably, the PDFs at all values of $\epsilon_\ell$ are very close to Gaussian, confirming the earlier observations and conclusions regarding KRSH. The only deviations from Gaussianity can be observed in the PDFs of longitudinal increments where the negative skewness for forward cascade and very slight positive skewness for inverse cascade conditioning are visible on either side of the peak of the PDFs.

\begin{figure} [H]
 \centering
  \includegraphics[scale=0.38]{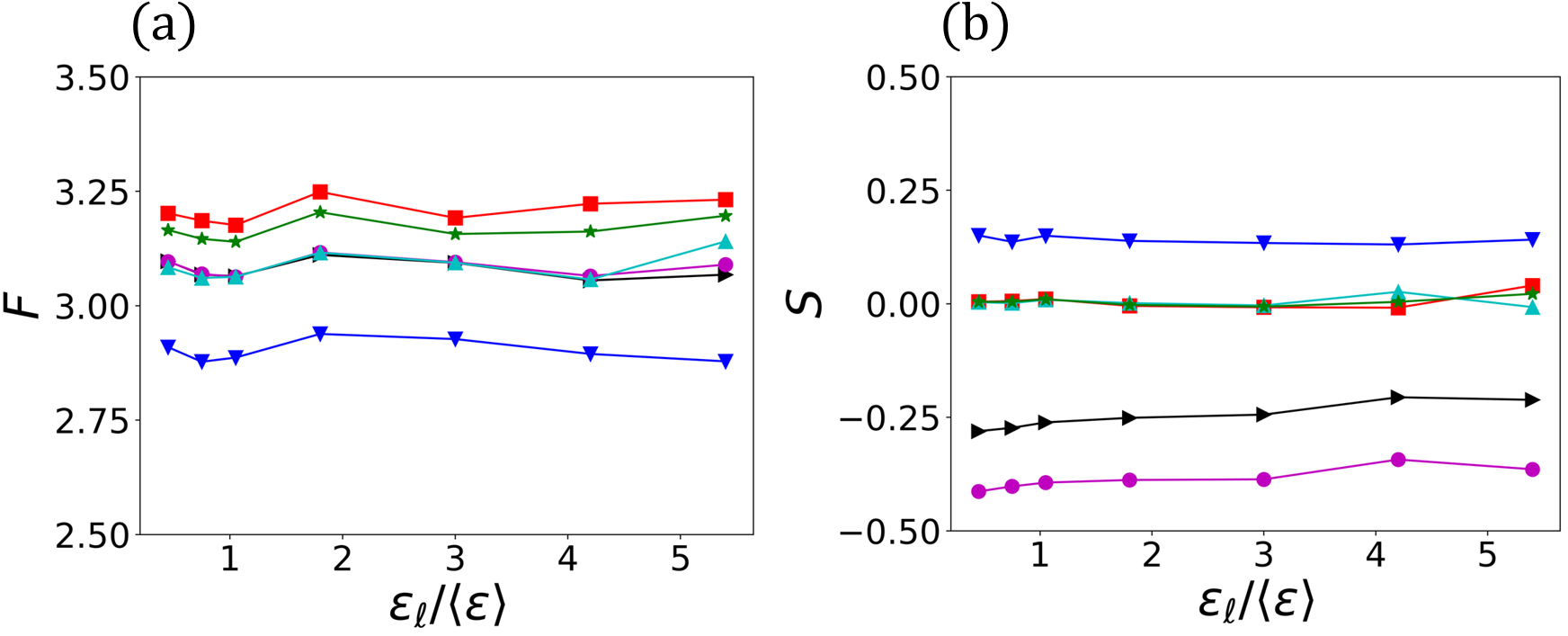}
    \caption{Panel (a) and (b) show flatness and skewness of velocity increments respectively. The notation of corresponding lines and symbols is identical to figure \ref{du2_45eta_eps}.} 
    \label{fig:FS_45eta_du_eps}
\end{figure}

The preceding observations about the PDFs are made quantitatively by measuring the flatness and skewness coefficients of the conditional PDFs. Results are shown in Fig. \ref{fig:FS_45eta_du_eps}. As is visible, the flatness coefficient is close to 3 (the Gaussian value) for all PDFs while the skewness coefficient is zero for the transverse components, and negative for global and forward cascade while it is positive (but smaller in magnitude than the negative) for inverse cascade regions. The values are relatively independent of the local rate of dissipation $\epsilon_\ell$.

\section{Filtered gradient statistics in forward and inverse cascade regions}
\label{sec:results2}
It is often argued that velocity increments at scale $\ell$ share many properties with elements of the velocity gradient tensor filtered at scale $\ell$ \citep{meneveau2011lagrangian}. We test this expectation in the present context of conditioning on the direction of the energy cascade. As before, we consider both longitudinal and transverse components of $\widetilde{A}_{ij}$, denoted as $\widetilde{A}_{ll}$ and $\widetilde{A}_{tt}$.

\begin{figure} [H]
 \centering
  \includegraphics[scale=0.5]{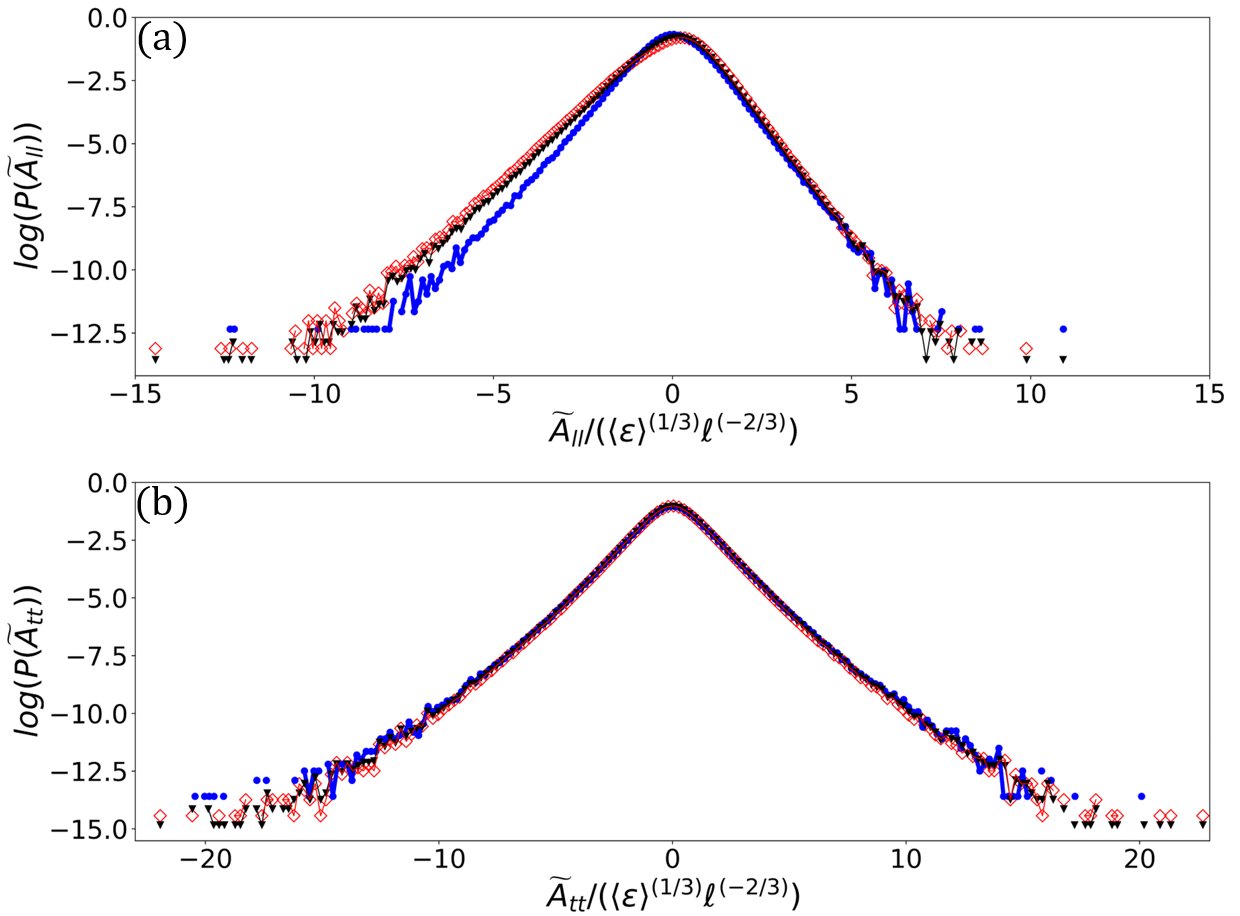}
    \caption{Panel (a): global (black triangle and line), and conditional PDFs of longitudinal filtered velocity gradient tensor at $45\eta$
    based on local forward (red open diamond and line) and inverse (blue circle and line) cascade rate. Panel (b): PDFs of transverse filtered velocity gradient tensor with the same notation as panel (a).}
    \label{fig: PDFs_45eta}
\end{figure}

A comparison of the results in Fig. \ref{fig: PDFs_45eta} with those in Fig. \ref{fig: PDFs_45eta_du} show that the trends of the PDFs for $A_{ll} $ and $A_{tt} $ under global and conditional averaging are at first sight similar to those of $\delta u_{l}$ and $\delta u_{t}$. The transverse components have a flatness near 6.3, independent of the direction of the energy cascade. Regarding the longitudinal component, the flatness of the global PDFs remains constant near 5, while the skewness is somewhat higher in magnitude compared to the skewness of $\delta u_l$, with values of $ S \sim -0.46$ compared to -0.29 (the larger values stem mainly from the smaller variance of the filtered gradients in the denominator). In the forward cascade regions, the skewness is again slightly more negative (-0.55). However, and contrary to the behavior of $\delta u_l$, the skewness of filtered longitudinal velocity gradient in inverse cascade regions is slightly negative ($S\sim -0.15$) as opposed to positive as one would expect from inverse energy cascade.

It bears recalling that the definition of $\Phi_\ell$ is based on the unfiltered velocity increments and so $\Phi_\ell<0$ is associated with positive velocity increment skewness, although the spherical averaging makes the connection not trivial. Conversely, filtering the velocity and then computing skewness yields a different measure of inverse cascade, perhaps relevant for scales in a range larger than $\ell$, while $\Phi_\ell$ is the transfer from scales smaller than $\ell$ to larger ones.

Since the negative sign of skewness of $\widetilde{A}_{ll}$ in inverse cascade regions is still somewhat counter intuitive, it could be due to the particular scale $\ell = 45 \eta$ chosen in the analysis. To test the robustness of the results shown above, we proceed to measure the PDFs of $\widetilde A_{ll}$ and $\widetilde A_{tt}$ conditioned on forward and inverse cascades at other three length scales within the inertial range, specifically $\ell={30, 60, 75}\eta$ (PDFs without conditioning are not presented here because they are similar to conditioning on forward cascade). The filtered velocity gradients are normalized using $\langle \epsilon \rangle^{(1/3)} \ell^{(-2/3)}$.

The excellent collapse of the PDFs at all scales considered confirms that the trends discussed above are quite robust. 

\begin{figure} [H]
 \centering
  \includegraphics[scale=0.5]{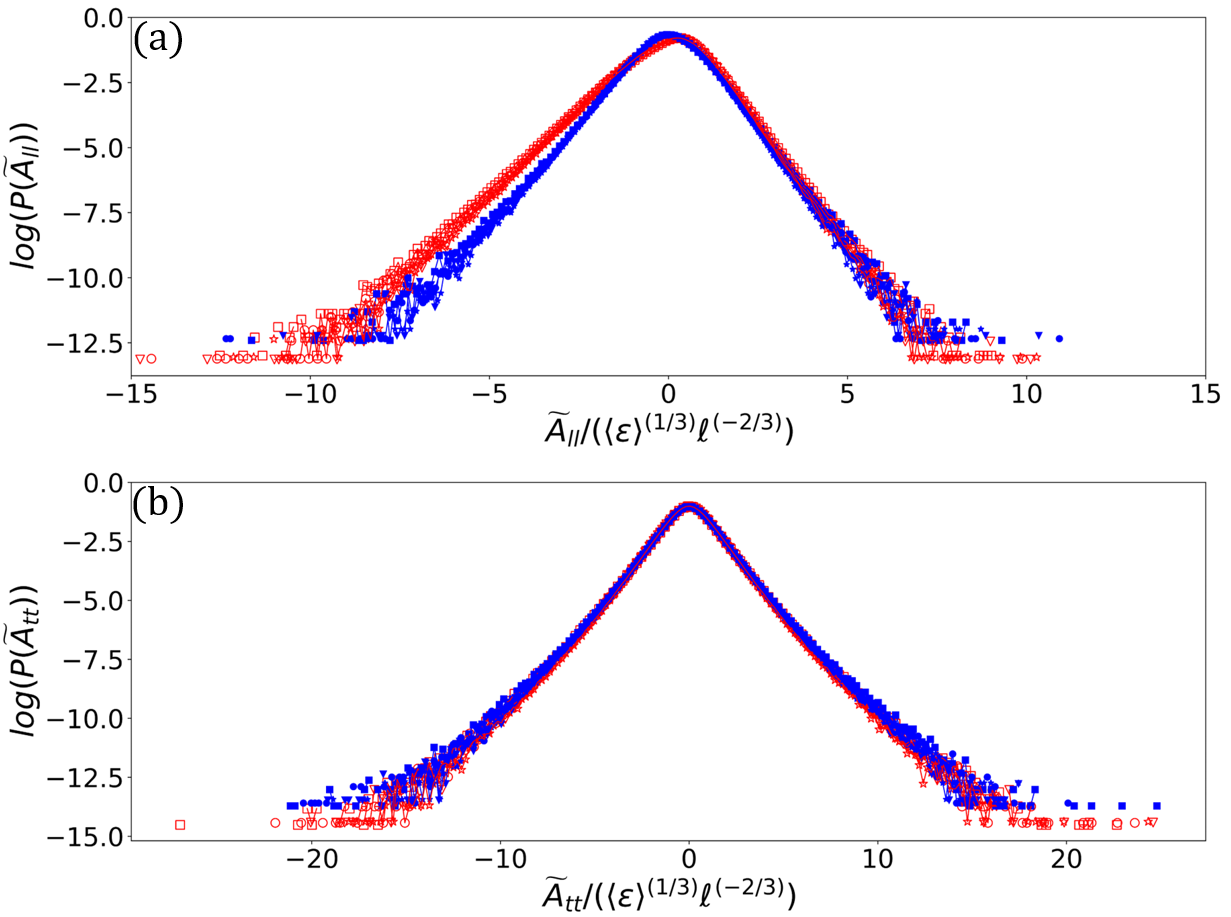}
    \caption{Panel (a): PDFs of longitudinal filtered velocity gradient tensor conditioning on local forward (red symbols and line) and inverse (blue symbols and line) cascade rates at $30, 45, 60, 75\eta$ (square, circle, triangle and star). Panel (b): PDFs of transverse filtered velocity gradient tensor with the same notation as panel (a).}
\end{figure}

\begin{figure} [H]
 \centering
  \includegraphics[scale=0.4]{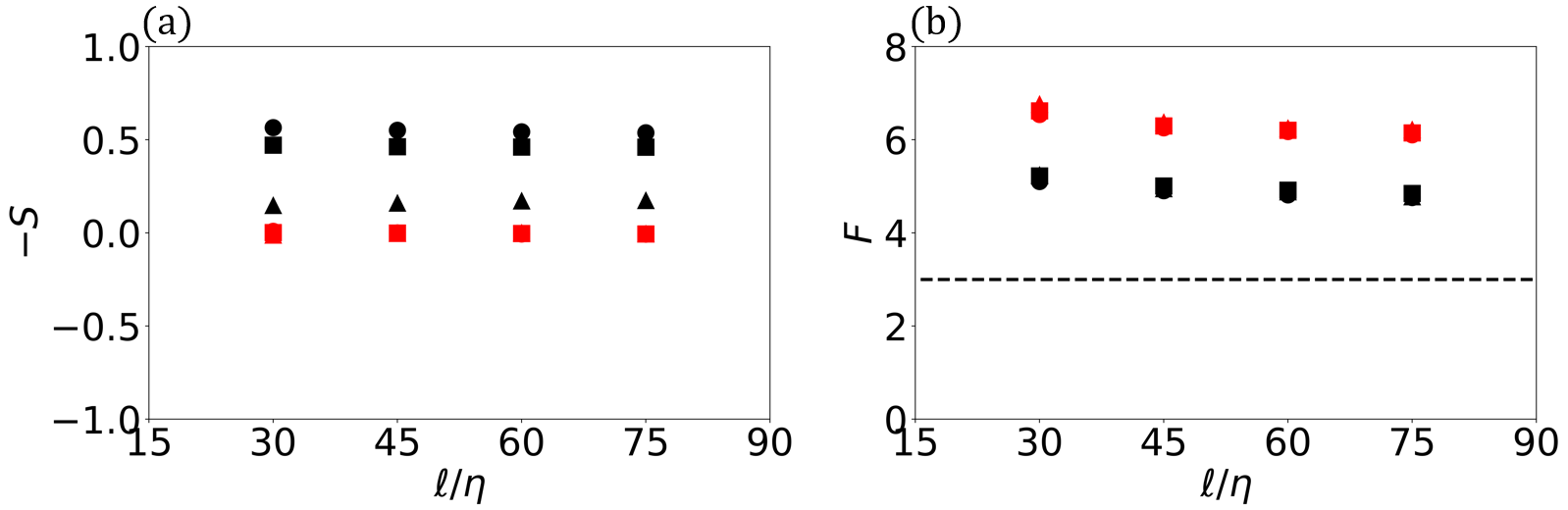}
    \caption{Panel (a): Skewness factor of longitudinal filtered velocity gradient tensor without conditioning (black square) and conditioning on forward (black circle) and inverse (black triangle) cascade rate at $\ell = 30, 45, 60, 75\eta$. The red squares are the results of transverse filtered velocity gradient tensor without conditioning. Panel (b): flatness with the same notation as panel (a).}
    \label{fig: SkandFla}
\end{figure}

In Figure \ref{fig: SkandFla}, we present the skewness   and flatness  factors for $\widetilde A_{ll}$ (depicted by black symbols) and $\widetilde A_{tt}$ (depicted by red symbols). For $\widetilde A_{ll}$, we display statistics for global averaging (squares), conditioning on forward cascade (circles), and conditioning on inverse cascade (triangles). It is evident that conditioning on inverse cascade consistently yields the lowest magnitude skewness, approximately $S \approx -0.16$, across all  scales considered. As for $A_{tt}$, $S \approx 0$ since the PDFs are symmetric. The flatness shows a slight increase with decreasing length-scale, as expected.  

\section{Conclusions}
\label{sec:conclusions}
We analyzed turbulence intermittency levels in regions of forward and inverse cascade, motivated by the possibility that inverse cascading regions could exhibit more noise-like Gaussian statistics in the inertial range of turbulence. The definition of forward and inverse local cascade regions was possible without conceptual ambiguity based on the scale-integrated KH equation. Levels of intermittency were quantified by measuring statistics of velocity increments and filtered velocity gradient tensor elements, both at scale $\ell$, the same scale across which the cascade rate was measured. A DNS dataset of isotropic turbulence at a moderately high Reynolds number was analyzed. The results show quite clearly that, contrary to the initial hypothesis motivating the present analysis, turbulence statistics remain highly intermittent in both forward and inverse cascading regions. 
Still, the forward cascade regions exhibit a more pronounced negative skewness, indicating slightly stronger non-Gaussianity than inverse cascade regions. When conditioning on local rate of dissipation, we find that KRSH holds for conditional velocity increments, equally in forward and inverse cascade regions. These results provide more detailed descriptions of the KRSH analysis presented in \cite{yao2024forward}. 

Conditional averaging identified a subtle difference between the skewness of PDFs of velocity increments of unfiltered velocity at scale $\ell$ and velocity gradient tensor elements filtered at the same scale. While the former exhibits a skewness consistent with the direction of the energy cascade, the latter displayed opposite trends, showing positive skewness also in inverse cascade regions. We can conclude that the physical quantity that represents inverse cascade rate ($\Phi_\ell$ when it is negative) depends sensitively on the inclusion of fluctuations of velocity at scales smaller than $\ell$ even if the velocity difference is evaluated over scale $\ell$. Perhaps unsurprisingly, these results suggest that resolved-scale information (at scales larger than $\ell$) contained in individual (longitudinal) elements of the filtered velocity gradient tensor is insufficient to properly describe the occurrence of inverse cascade. However, we recall that the results of \cite{yao2024comparing} demonstrated that the full tensorial invariants of the velocity gradient tensor (joint statistics with the $(R,Q)$ invariants) did contain  information about the presence of inverse cascading. In conclusion, while the single-component filtered gradient statistics proved insufficient to predict inverse cascading, the information contained at large scales via the joint tensor invariant statistics (depending on all tensor elements) \citep{yao2024comparing} appears more appropriate to detect inverse cascade regions based on large-scale information. 

It is important to stress that the present results, such as negative skewness of the longitudinal components of filtered velocity gradient conditioned upon flux direction depend upon the representation of flux chosen in this work, i.e. $\Phi_\ell$ based on third-order structure function with local spherical averaging.  If instead we were to use the LES formalism to define energy flux based on the subgrid stress tensor (see \cite{yao2024comparing} for detailed comparisons), one would expect better correspondence between the skewness and the sign of the local cascade. A better correspondence is expected since it is known that the nonlinear model \citep{borue1998local} using a spatial local filter (such as the box filter) yields energy fluxes that are well correlated with the flux defined in the LES sense using filtering (e.g. box filtering). This includes negative fluxes. Our focus on $\Phi_\ell$, i.e. the flux definition based on structure functions, stems from the fact that it corresponds locally to a flux in scale space (divergence with respect to ${\bf r}$ rather than simply a sink/source term at a fixed scale as in the LES formalism), i.e., it has a more unambiguous physical interpretation as a flux across scales.

The present data analysis was greatly facilitated by an open database system containing large amounts of turbulence data.   The refactored Johns Hopkins Turbulence Database (JHTDB v2.0) and its improved data access methods ({\it getData} and {\it getCutout}) are built around a portable Python package, which provides a more efficient, scalable, and user-friendly open source framework for accessing and analyzing turbulence data.  The refactored system is introduced and described in some detail in the appendix.

\section*{Acknowledgements}
The help from the IDIES technical team is gratefully acknowledged. 

\section*{Funding}
Funding for this project is provided by the National Science Foundation (Grant \# CSSI-2103874. Storage was supported by a CC* grant from NSF (\# 2322201). 

\section*{Appendix A: The refactored JHTDB system}

Ever since its inception \citep{li2008public}, the public turbulence database system (Johns Hopkins Turbulence Databases, JHTDB) has provided datasets for many scientists and projects, leading to over 400 peer-reviewed papers in the turbulence literature (see publication listings at {\it https://turbulence.idies.jhu.edu/publications}). The original system was based on DNS data that was stored in an SQL database system, with a Z-order fractal space-filling curve used as an indexing method. Unlike the traditional IJK indexing, this data indexing had the advantage that points located close by in 3D would be (with high probability) also located close by along the space-filling curve and index. This helped increase access speeds for sequential disk reads. Small $8^3$ data voxels were used as the elemental data unit. Operations such as interpolation and differentiation were done close to the data using pre-programmed instructions in User Defined Functions written in C$\#$. Users could access the data using Matlab, Fortran, C and python which interfaced with the SQL database system using Web services based on the SOAP interface. Following the initial implementation for several spatially homogeneous turbulent flows, several improvements were implemented and reported in this journal, such as Lagrangian fluid particle tracking capability \citep{yu2012studying} and non-homogeneous fully developed channel flow data \citep{graham2016web}. In the meantime, however, file systems, programming languages, and storage hardware have evolved significantly. Also, the SQL-based system (a commercial software suite) was difficult to replicate elsewhere.  To address the need for more open science methods, the JHTDB has been refurbished completely by porting data analysis codes from C\# running in the legacy JHTDB SQL Server system to Python running on the servers close to the data, using as much as possible distributed/parallel frameworks such as multi-threading. Below we provide an overview of the main elements of the refurbished system. As of this writing, some of the old existing datasets are still based on, and being served out of, the old system, but they are being transitioned to the new system. These changes are not visible to users since the new data access tools are built to be (mostly) backward compatible with the old system. 

\subsection*{A1: Data storage using Zarr format}\label{arch}
We develop a method to store data using chunked Zarr files (a scalable, compressible, and versatile array storage format (https://zarr.dev/)) rather than the Z-ordered approach formerly used for JHTDB datasets. This pivot to Zarr files resulted in several benefits. First, we were able to make the processing code much simpler and thus easier to maintain and share. Second, processing times became significantly faster with Zarr files due to fewer distinct I/O operations when reading data. Third, converting to Zarr files is enabling us to more easily transition to Ceph, an object storage system. New datasets are stored on a Ceph cluster. 

\begin{figure}
 \centering
  \includegraphics[scale=0.35]{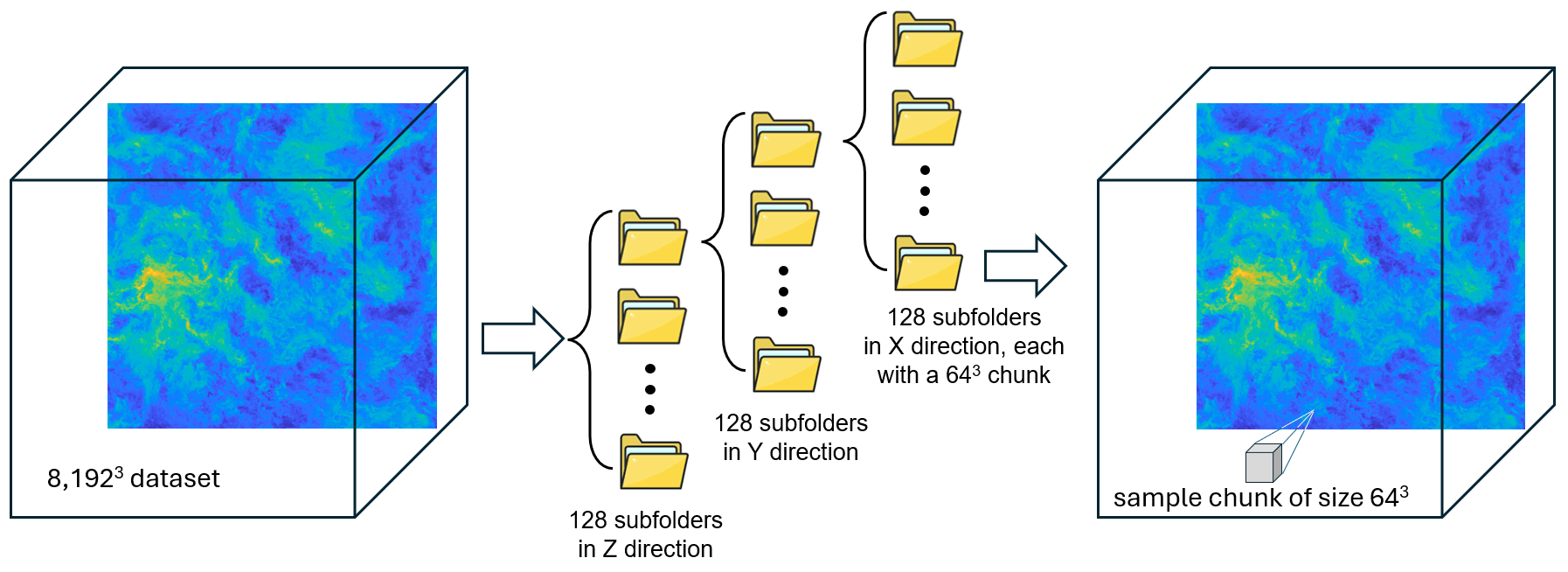}
    \caption{Schematic diagram of a $8{,}192^3$ data cube stored in Zarr format on the Ceph cluster. Chunks of $64^3$ points have been found to be an optimal size to facilitate rapid access over a variety of most common expected access patterns.}
    \label{fig:zarr}
\end{figure}

The Zarr file storage approach is illustrated based on the $8{,}192^3$ DNS dataset used in the analysis presented in this paper. A given $8{,}192^3$ cube of data is stored as a single Zarr store with chunked data cubes of size $64^3$. We have experimented with many options and this size has been found to be optimal with expected typical usage patterns. Inside the Zarr store, there are therefore $128^3 = 2{,}097{,}152$ folders ordered in KJI indexing, each containing a $64^3$ chunk of data (see Fig. \ref{fig:zarr}).

\subsection*{A2: Turbulence services using python notebooks: the GetData function}
\label{getdata} 
A new Python package ({\it ``Giverny''}) has been developed, which forms the backbone of the new data access method used by the refactored JHTDB system. Users call a single data access function, ``{\it getData}''. 
 The new {\it getData} function communicates with new datasets (e.g. the $8{,}192^3$ isotropic turbulence and stably stratified atmospheric boundary layer datasets) on the Ceph cluster through python-based source code, while any other legacy datasets on SQL cluster are accessed through the Python SOAP interface. The Python SOAP interface is included in the getData function, so users do not need to install it. In this way, the new approach unifies previous getfunctions, including fields, gradients, Hessians, and Laplacians of velocities and pressure, with existing spatial and temporal interpolation methods, into one general {\it getData} function. The user does not have to specify the underlying file system, which is accessed automatically depending on the dataset chosen and whether that dataset has already been transferred to Ceph or not.  This approach enables the gradual transition of all datasets to the new Zarr format. We provide two ways to access the getData function using Python: (1) On SciServer, where the source code is directly executed close to the data. (2) Python Jupiter Notebook that can be executed on a  user's own local computer, using REST web service interfacing to the database. In both cases, users can pip install the required `{\it Giverny}' Python packages, called `{\it Giverny}' and `{\it Givernylocal}', respectively.  Error messaging is implemented where user input is checked after {\it getData} is called and informative feedback is reported to the user if any parameters are not specified correctly.

\begin{figure}
 \centering
  \includegraphics[scale=0.18]{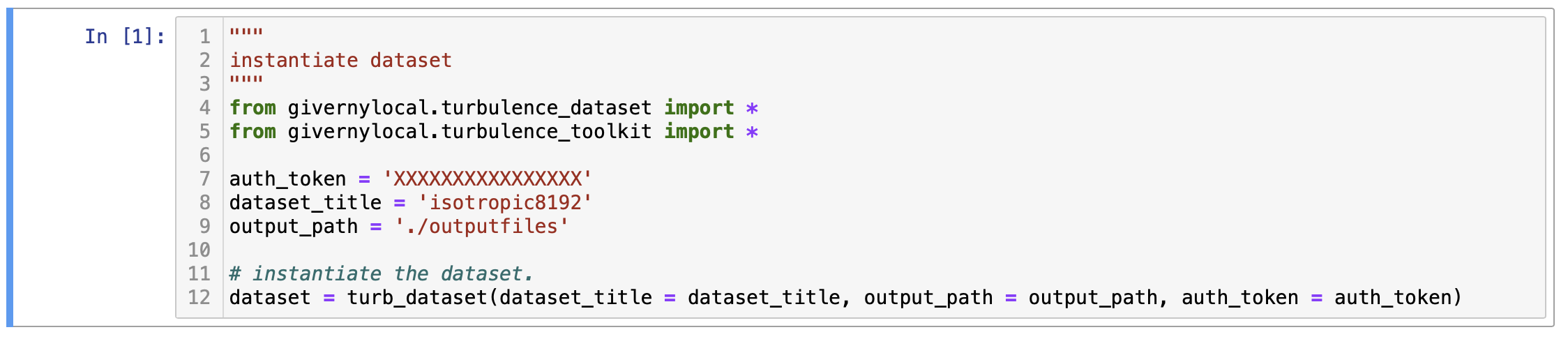}
    \caption{Snippet of Python DEMO notebook, instantiating the dataset, where users load {\it Giverny} package, set authorization token and specify dataset name.}
    \label{fig:2D_demo_ini1}
\end{figure}

\begin{figure}
 \centering
  \includegraphics[scale=0.19]{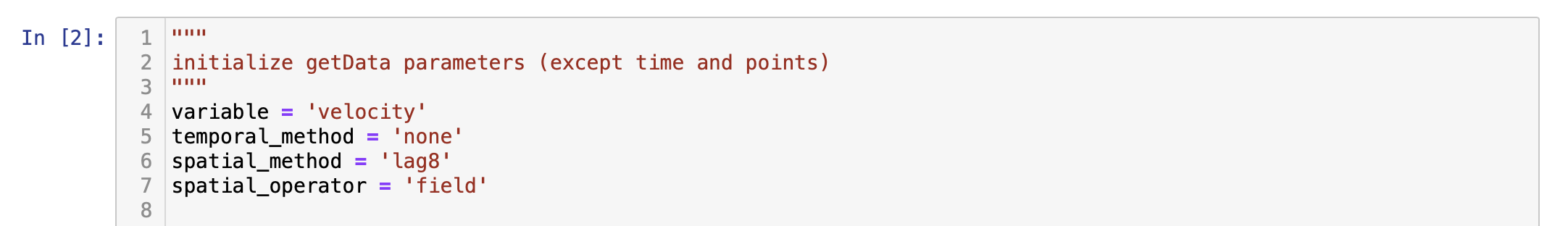}
    \caption{Snippet of Python DEMO notebook initializing {\it getData} parameters such as interpolation methods (available options shown in figure \ref{fig:2D_demo_inter}.}
    \label{fig:2D_demo_ini2}
\end{figure}

In order to illustrate the usage of the {\it getData} function, an example from a DEMO Python code available to users is reproduced below.  The instantiation of the dataset and initialization of parameters is demonstrated in Figures \ref{fig:2D_demo_ini1} and \ref{fig:2D_demo_ini2}.

\begin{figure}
 \centering
  \includegraphics[scale=0.22]{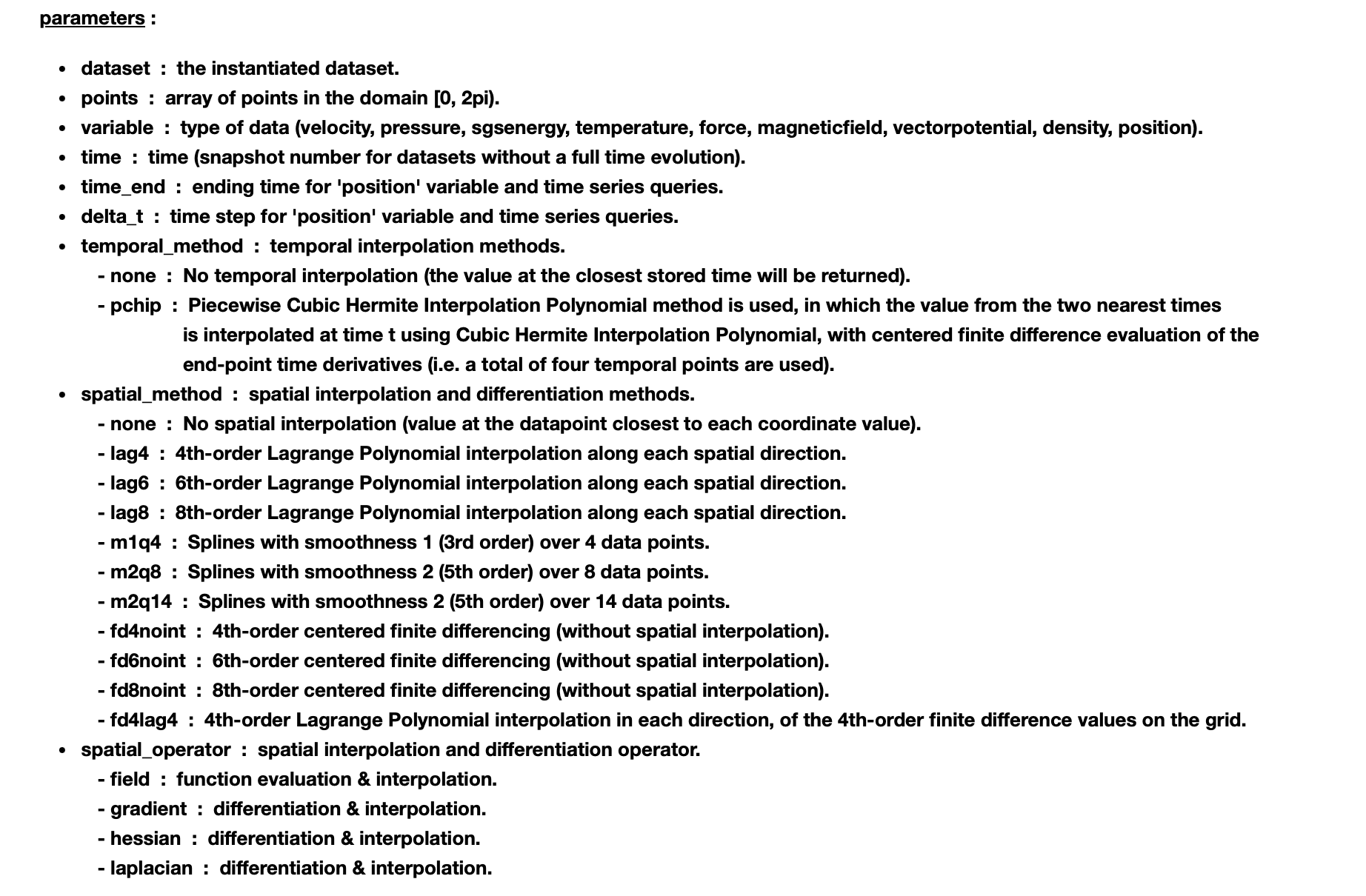}
    \caption{Description of input parameters for {\it getData} function such as  spatial and temporal interpolation methods.}
    \label{fig:2D_demo_inter}
\end{figure}

The DEMO code  requesting a 2D plane array from the $8{,}192^3$ dataset used in this paper is shown below. Users can specify the name of datasets (both legacy and new turbulence datasets), variables (e.g. velocities, pressure, temperature, magnetic field, etc.), time points (time steps), temporal/spatial interpolation methods (listed in the first cell of the demo code), spatial operator (Field, Gradient, Hessian, Laplacian), points array (any requested 3D coordinates), and options (any other functions for specific turbulence datasets, for example, getPosition for isotropic1024 and channel flow). Figure \ref{fig:2D_demo_inter} displays a list of available options.

\begin{figure}
 \centering
  \includegraphics[scale=0.2]{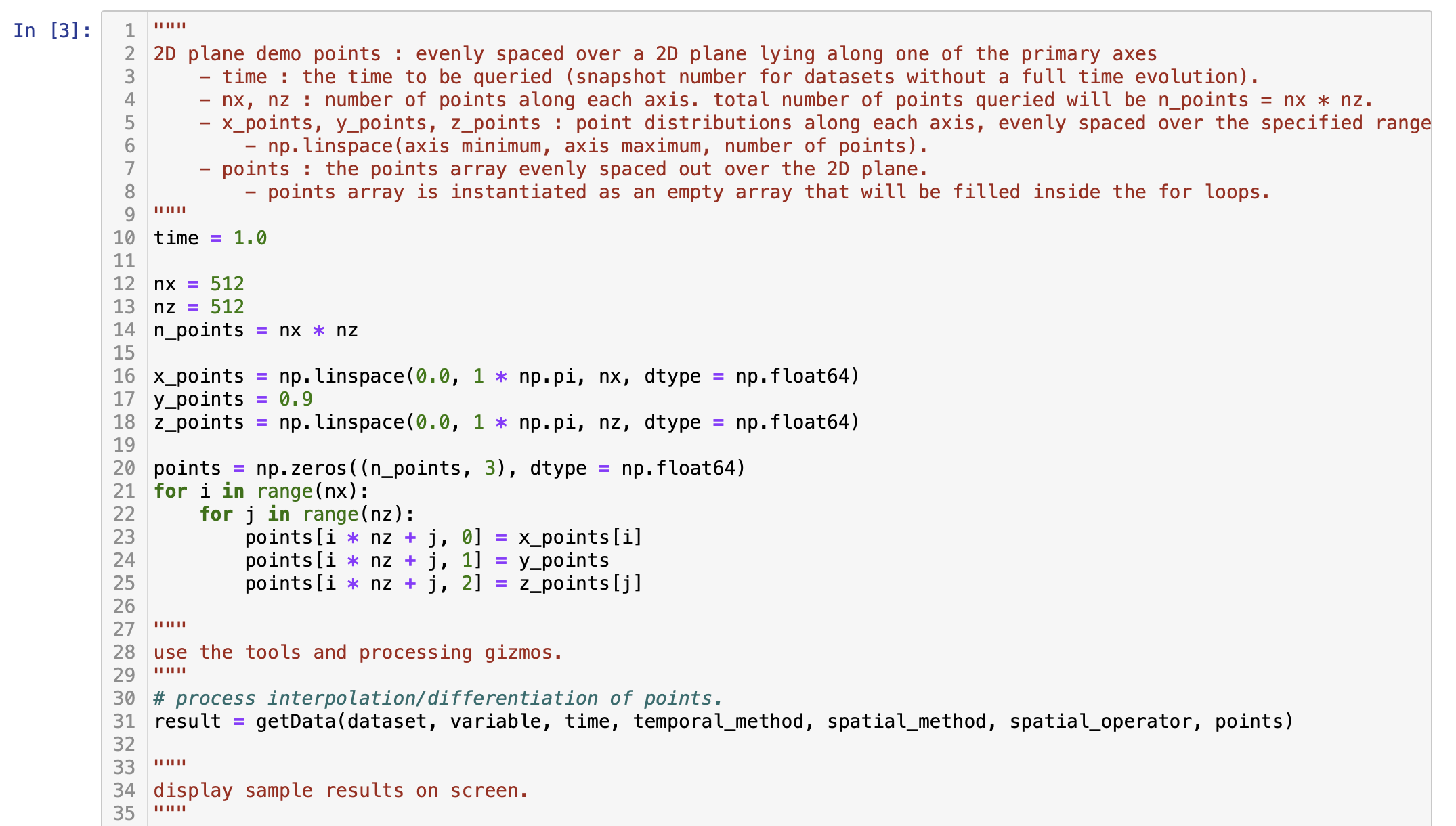}
    \caption{Python demo code of using getData approach, requesting points on a 2D plane of $512\times 512$ equispaced points on an $x-z$ plane between $0-\pi$ and at $y=0.9$ .}
    \label{fig:2D_demo}
\end{figure}

\begin{figure}
 \centering
  \includegraphics[scale=0.17]{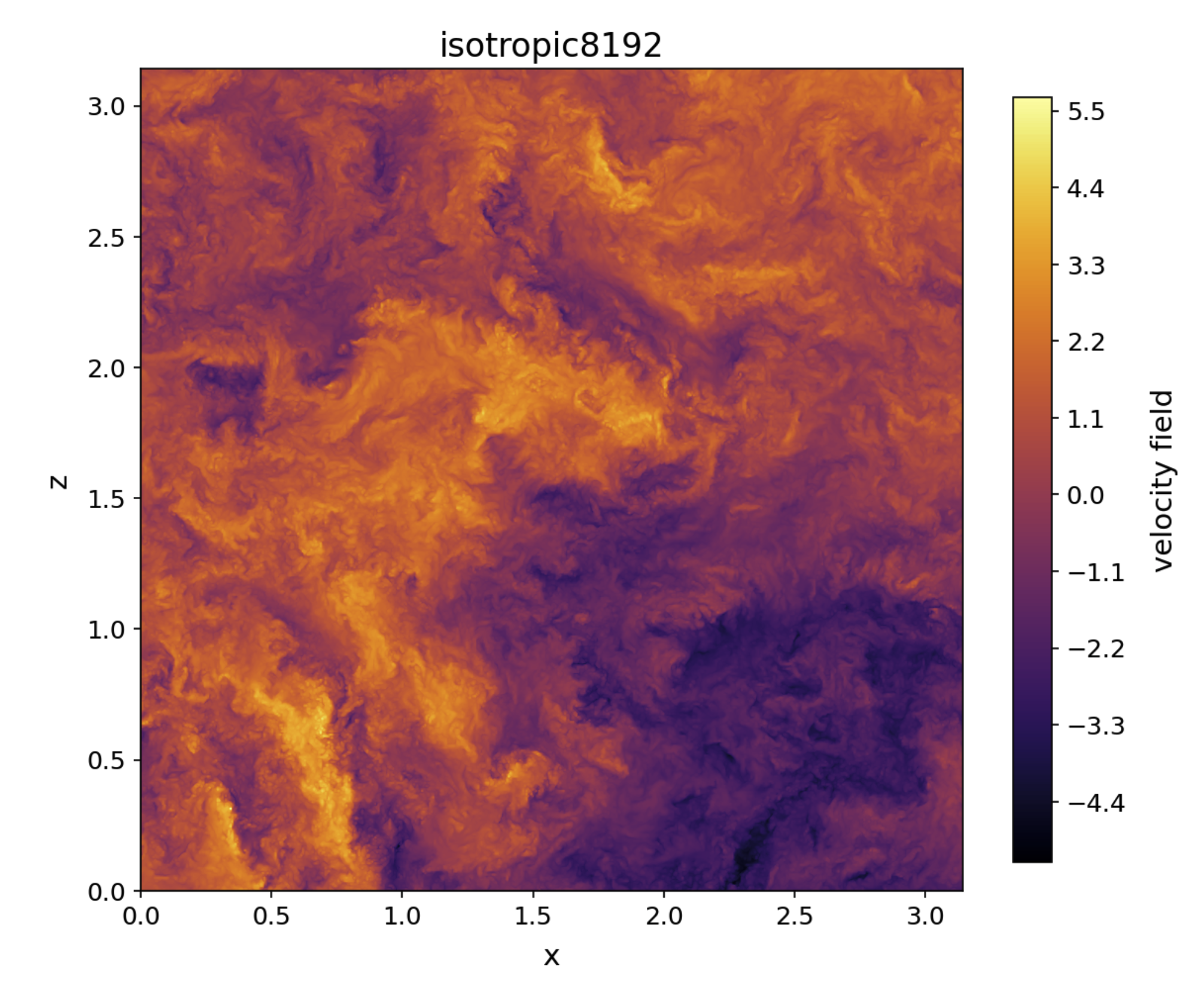}
    \caption{Contour map of velocity on $512 \times 512$ points distributed on a 2D plane of the y-component velocity   at $y=0.9$ as obtained from executing the code shown in Fig. \ref{fig:2D_demo} above.}
    \label{fig:2D_demo_contour}
\end{figure}

In this example, the array of points at which data are requested is located on a plane. They can equally be located on a cube, randomly distributed, or data on a point but as a function of time can be requested. The DEMO notebook provided to users illustrates such data access modes. 

\subsection*{A3: Turbulence services using python notebooks:  The   getCutout function}\label{cutout}

\begin{figure}
\centering
\includegraphics[scale=0.20]{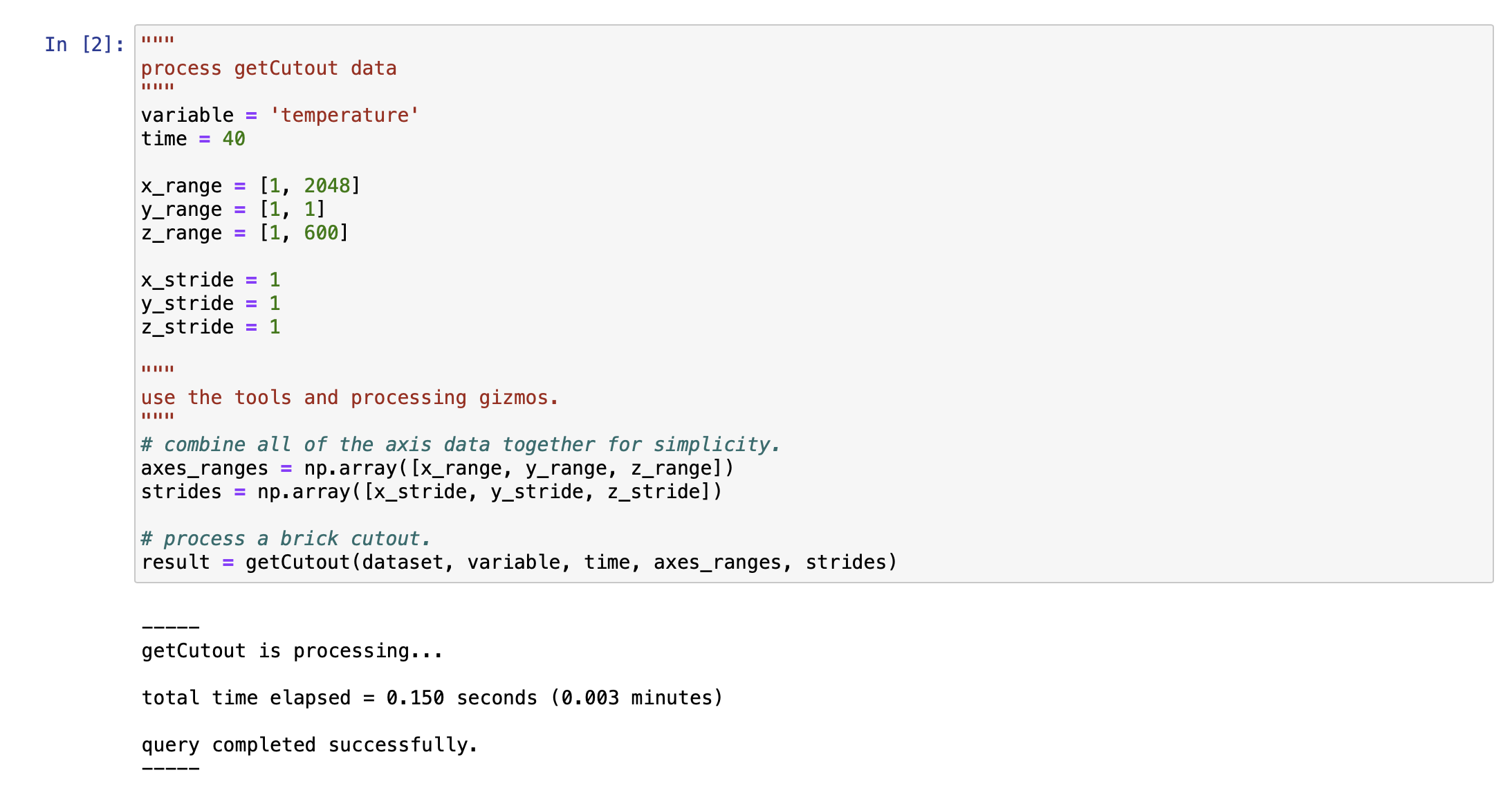}
\caption{\label{fig:getcutout_processing_code2}Python demo code for {\it getCutout} selecting a $x-z$ plane cutout for requesting a rectangular cutout of temperature data.}
\end{figure}

An updated {\it GetCutout} notebook has also been developed as part of the refactored JHTDB methods. This notebook has to be executed on SciServer, close to the data.  Unlike {\it getData} where users specify times and locations in actual physical variable values (e.g. $(x,y,z,t)$) that do not need to coincide with discrete data points, in {\it GetCutout} user specifies integer index values for position and time steps. Data is returned as arrays on the specified set of grid points.  User input is checked before query submission and informative feedback is reported to the user if any getCutout input parameters are not specified correctly. A Python demo notebook with the functionality to query all JHTDB datasets, and their associated variables, is illustrated below by means of a notebook snippet. 
Figure \ref{fig:getcutout_processing_code2} shows the specification of the time-step (40) and ranges of grid-points in $x$ (horizontal) and $z$ (vertical) directions. Execution on SciServer is very fast, 0.15 seconds in this case to extract temperature at over 1.2 million points.  
Once a query is finished processing, the user can choose to plot directly in the notebook or save the cutout to HDF5 and XMF files for further analysis outside the notebook.  

\begin{figure}
\centering
\includegraphics[scale=0.23]{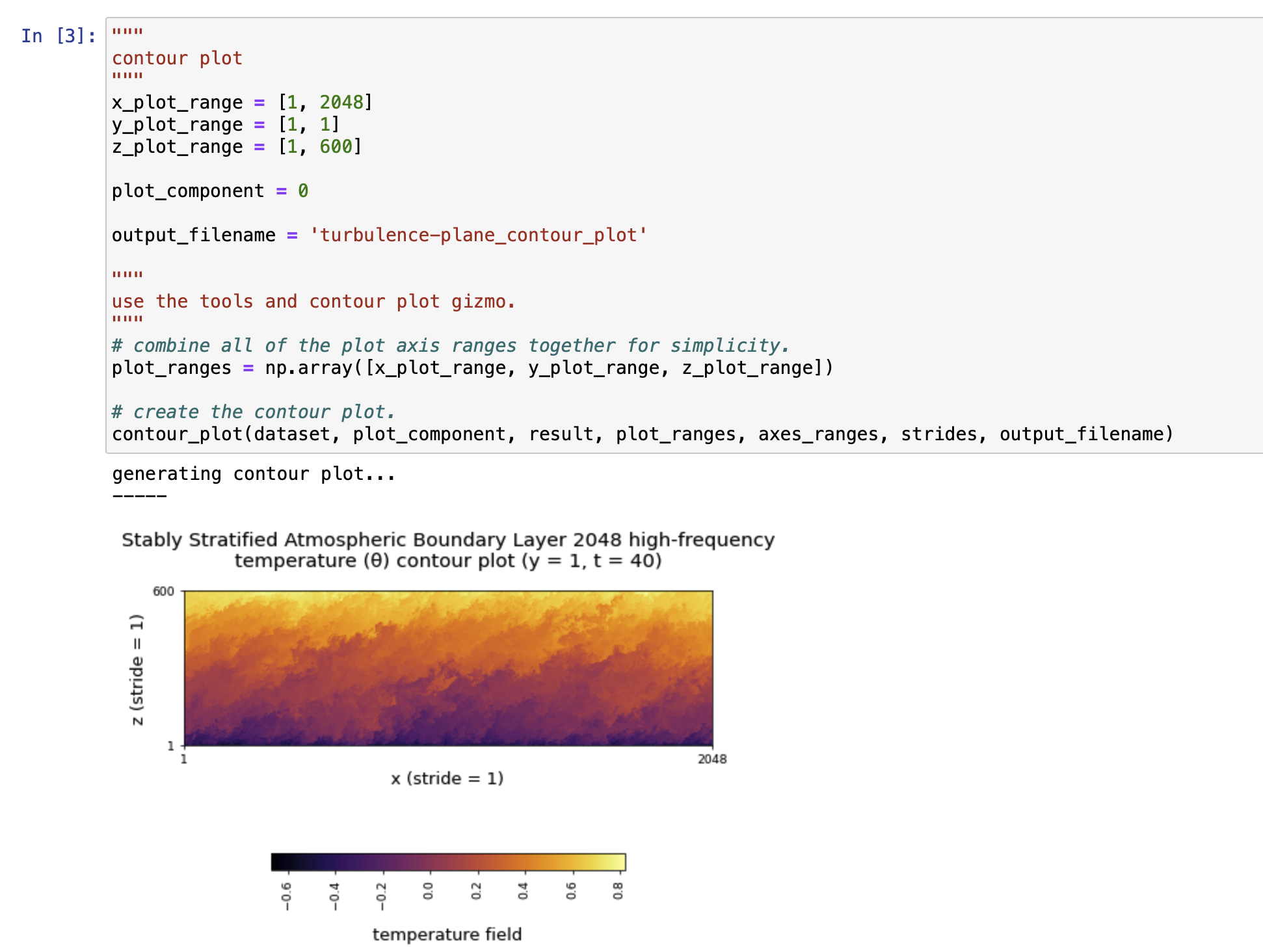}
\caption{\label{fig:getcutout_plot}Python demo code plotting the requested temperature (deviation from reference temperature) as function of grid points. }
\end{figure}

The temperature contour plot from the requested cutout of the new Stable Atmospheric Boundary Layer (SABL) dataset is shown in Figure \ref{fig:getcutout_plot}. The temperature is lower near the ground and the characteristic ramp-cliff structures are clearly visible. Requesting such planes at successive time steps allows for the generation of animations, etc.

\begin{table}[!htb]
    \small 
    \setlength{\tabcolsep}{4pt} 
    \begin{center}
    \renewcommand{\arraystretch}{1.3} 
        \begin{tabular}{|l||c|c|c|}
            \hline
            \multicolumn{1}{|p{2.8cm}||}{\centering\textbf{Dataset}}
            &\multicolumn{1}{|p{2.8cm}|}{\centering\textbf{Cutout shape}}
            &\multicolumn{1}{|p{2.3cm}|}{\centering\textbf{Files\\(\#)}}
            &\multicolumn{1}{|p{2.3cm}|}{\centering\textbf{Time\\(seconds)}}\\
            \hline \hline
            isotropic 1024 & $1024^2$ & 4 & 3\\
            \hline
            channel flow & $256^3$ & 2 & 15\\
            \hline
            transitional boundary layer & $224^3$ & 1 & 12\\
            \hline
            isotropic 8192 & $8192^2$ & 1 & 30\\
            \hline
            stable atmospheric boundary layer & $2048^2$ & 1 & 2\\
            \hline
        \end{tabular}
    \end{center}
    \caption{Timings of velocity cutouts for various datasets.}
    \label{table:getcutout_timings}
\end{table}

Example {\it getCutout} timings for several datasets are shown in Table \ref{table:getcutout_timings}. 

\subsection*{A4: {GetData} turbulence services using Matlab, Fortran, and C}\label{MatlabFC}

JHTDB also provides DEMO codes in Matlab (not shown) which calls a Matlab function {\it GetData.m} which then accesses the database  through the REST service. It communicates with the Python {\it getData} function executing on a Kubernetes web cluster near the data. The process of instantiating the dataset and initializing getData parameters is similar to the Python method.
(see Figures \ref{fig:2D_demo_ini1} and \ref{fig:2D_demo_ini2}). 
The getData.m file is included in the Matlab package. It includes a single URL that can access the new and legacy datasets,   replacing the previous cumbersome Matlab SOAP interface files that were required for each of the many old {\it get} functions. Access functions in C and Fortran are similarly available, also communicating with the data through the REST service and the Curl and Fortran-Curl packages, respectively.

\vfill

\newpage 

\bibliographystyle{apacite}
\bibliography{interactapasample}

\end{document}